\documentclass[showpacs,twocolumn,amsmath,superscriptaddress,floatfix,pre]{revtex4}
\usepackage{epsfig,verbatim}

\begin{document}

%
\title{Statistics of Microstructure Formation in Martensitic
Transitions  Studied  by  a  Random-Field Potts  Model  with  Dipolar-like Interactions}
\author{Benedetta Cerruti}
\email{benedetta@ecm.ub.es}
\author{Eduard Vives}
\affiliation{ Departament d'Estructura i Constituents de la Mat\`eria,
  Universitat  de Barcelona  \\ Mart\'i  i Franqu\`es  1,  Facultat de
  F\'{\i}sica, 08028 Barcelona, Catalonia, Spain}
\date{\today}
\begin{abstract}
We have developed a  simple model  for the  study of  a cubic  to tetragonal
martensitic  transition, under athermal  conditions, in  systems  with  a
certain  amount of  disorder.  We have performed numerical  simulations that
allow for a statistical study  of the dynamics of the transition when
the  system is driven  from the  high-temperature  cubic phase  to the
low-temperature degenerate tetragonal phase. Our goal is to reveal
the existence of kinetic  constraints that arise from competition
between  the equivalent  variants of  the martensitic  phase, and which
prevent the system from reaching optimal final microstructures.
\end{abstract}

\maketitle

\section{Introduction}
\label{Introduction}
The kinetics  of thermoelastic martensitic transitions  (MT) 
is a  challenging problem to which physicists  and material scientists
have  devoted  several  decades  of  study. The  complexity  of  the
phenomenon arises  from three main factors:  (i) Firstly,  the
first-order character  of the transition which, in  many cases, implies
the  existence  of  high-energy barriers  and  associated  long-lived
metastable  states.  The  word "athermal"  has been  used to  refer to
those MT  where thermal  fluctuations do not  play any role,  and thus
only evolve through states  when metastability limits are reached.
This, combined with the existence of disorder in the systems, leads to
history  dependence and avalanche  dynamics. (ii) The  second important
factor  is the  non-trivial  relation  between the  symmetries of  the
parent and  product phases, especially in real  3D systems.  
The  product phase
may appear in a number of energetically equivalent variants which have
the tendency to satisfy some "matching" conditions between each other and 
the  parent  phase.   (iii)  The  third  contribution to  the
complexity  of   the  phenomenon  is  the  existence   of  long-range
correlations due to the elastic nature of the problem.  The effects of
nucleation of a martensite domain in a certain point of the system 
may influence
the kinetics  of the transition at a large distance from this  point.  The final
microstructures observed  in the martensitic phase  are a consequence
of the  complex interplay of these  three factors (i-iii),  and thus 
if one wants to have a full understanding of the phenomenon,
it is not enough to reduce it to the problem of the optimization of a 
certain Hamiltonian. In order to understand 
this point of view, it is  instructive to
compare real microstructures observed  over a large sample (see, for
instance  Fig.1 in  Ref.~\cite{Zhang2008} and Fig.4 in Ref.~\cite{Cui2004}) 
with  the  perfect patterns
calculated    from   elasticity    theories    (see,   for    instance
Ref.~\cite{Bhattacharya2003}).  The definition of  "kinetic constraint"
in much simpler models has
been introduced  \cite{Toninelli2006,Toninelli2006b} 
to account for the  fact that some  low-energy states
cannot  be obtained  due to  the  absence of  a possible  pathway.
Current   theories  are  still   far  from   being  able   to  predict
martensitic microstructures.   
Details such as sample  size  and  shape,  defects,
annealing  times,   quenching  rates,   etc.,  are  known to dramatically change
the final state of  the martensite 
(similarly to what happens for the domain
structure of magnetic materials \cite{schaefer,barkhausen}).  In fact, it has been known
for centuries  that to obtain a "good" microstructure  in a metallic
alloy is an "art" reserved for the best smiths or metallurgists.

Quantitative experimental information concerning the
growth    of   the    microstructure   in    martensitic   
transitions is difficult to be obtained
\cite{Vantendeloo1997}.   This  is   because  most
observation techniques,  both direct imaging  or scattering techniques
(X-rays, neutrons and electrons microscopy), cannot be easily 
performed ``in situ'' during the
transition. One must  stop the transition in an  intermediate state by
quenching   and  perform   a  posterior   sample  treatment   for  the
observation. Furthermore, 
in  the case of direct imaging  techniques it is
difficult to  extract numerical data from sequences  of 2D micrographs
of the  system surface. There  have been some ``in  situ'' synchrotron
radiation experiments \cite{Mitsuka2006} that follow the growth of
peaks associated with martensitic structures, but  the studies have
mostly concentrated on understanding  the precursor effects before the
real transition starts.  Some interesting phenomena have been reported
during  the transition. In  some cases  the formation  of intermediate
``phases''  with only  short-range order  and  with reciprocal  space
vectors,  different from the  final position  of the  martensitic peak
structure have been observed \cite{Vantendeloo1997}. In addition, 
a recent work has
quantified   the   fact   that   microstructures   exhibit   imperfect
self-accommodation   due   to   the   coalescence   of   the   variants
\cite{Zhang2008}.
  
Among the  theoretical efforts to  understand microstructure formation,
one  should mention  many  continuum models,  derived from  elasticity
theory       \cite{Wang1997,Ichitsubo2000,Ahluwalia2001,Jacobs2003,Hatch2003,Lookman2003,Rasmussen2001}.   Despite advances in  computing
power,  the complexity  of the  models (especially the  long-range
character of the interactions) is such that it is difficult to perform
a  large  number  of   simulations  and  a statistical  analysis  of
microstructures. In many cases the models have only been applied to 2D
unrealistic crystals, while in others the number  of coexisting variants
has been  reduced artificially or  by considering a  symmetry breaking
external stress.

In this work  we will focus on the cubic to  tetragonal MT. This kind of
transition occurs,
for  instance, in  a number  of binary  metallic alloys  for different
ranges  of concentrations.  Among  others \cite{Bhattacharya2003} 
we recall InTl,
InPb,   NiAl,    FePt   and   FePd    \cite{Mitsuka2006}.    Different
microstructures  have been  observed  in these  transitions.  A  common
basic feature is  the clear tendency for the  variants to grow forming
twins.  These  are domains of  different variants that share  a common
interface  oriented  in  a   certain  preferred  direction.   By  this
mechanism, elastic distortions of  the lattice are minimized.  For the
cubic to  tetragonal transition of interest here,  the twin boundaries
are planes perpendicular to the (110) direction  (or
any  of   the  6   equivalent  ones) of the cubic phase. 
The   overall  microstructures,
nevertheless, are a combination of the different twins and usually look
quite random.

In  a previous paper  we introduced  a Random  Field Potts  Model with
truncated  dipolar   interaction  for  "athermal"   first-order  phase
transitions between phases with any  change of symmetry \cite{Cerruti2008}. The model was
based on the $T=0$ RFIM with metastable dynamics, which was originally
proposed for  the study of field induced  athermal transitions between
two  ferromagnetic symmetrically  equivalent phases. The  proposed
modifications  allowed,  for  instance, the following to be studied: transitions from  a
non-degenerate phase  to a phase  with three different  variants, the
influence  of the  amount  of disorder  on  the dynamics  and how  the
parameters  of   the  dipolar   interaction  term  affect   the  final
microstructure.

In the present study we are  going to go one step forward. 
We will adapt the
model to the  symmetries of a cubic to tetragonal  MT.  We will extend
the  range of  the dipolar  interaction in  order to  obtain  the twin
microstructures observed in  such a MT transition. We  have found that
it is enough to include up to 4th nearest neighbors. 
By truncating long-range
interactions we accelerate  our simulations, so that we  can perform a
systematic  study of  the Hamiltonian  parameters and  statistics over
disorder configurations.

The paper is organized as follows: in Section \ref{model} we introduce
the  Hamiltonian of  our  model,  and discuss  the  truncation of  the
dipolar  term.   We also  give  details  about  the dynamics  and  the
simulations.   In  Sec.\ref{groundstate}  we  present a  ground  state
analysis of the martensitic  microstructures that optimize the energy.
This will  help us  in choosing the  interesting range  of Hamiltonian
parameters.   In   Sec.   \ref{simmicro}   we  present  some   of  the
microstructures obtained  by dynamically  driving the system  from the
parent  to the  fully saturated  martensitic phase.   The  results are
compared   with   the   previous  ground-state   configurations.    In
Sec.\ref{isteresi} we  discuss the shape of the  hysteresis cycles and
the effect of varying the Hamiltonian parameters (disorder and dipolar
intensity) and the  sample size.  A more quantitative  analysis of the
microstructures  is  presented  in  Sec.\ref{real_space_analysis}  and
\ref{fourier_analysis}, were  we show  the system evolution  along the
hysteresis cycle in real and Fourier  space, respectively.  Finally in
Sec. \ref{Conclusions} we both summarize and conclude.

\section{model}
\label{model}
Let us consider  a simple cubic lattice of  size $N=L\times L\times L$,
with lattice parameter $a$ and 
periodic  boundary conditions. At  each lattice site we  define a
variable which can  take four different values that  we will call $\hat{0}$,
$\hat{x}$,  $\hat{y}$ and  $\hat{z}$. 
As  in  our previous  work \cite{Cerruti2008},  we
have chosen to represent our  variables by considering a vector $\vec{S_i}$
($i=1,\dots N$),  having three components:  we will indicate  the four
possible     values    as     $\hat{0}=(0,0,0)$,    $\hat{x}=(1,0,0)$,
$\hat{y}=(0,1,0)$  and  $\hat{z}=(0,0,1)$.   These  variables  can  be
interpreted as  an elementary domain in the cubic  austenite phase (vector
$\hat{0}$),  or  in  the  three  possible  variants  of  a  tetragonal
martensitic   phase  ($\hat{x}$,   $\hat{y}$   and  $\hat{z}$).    The
interaction  between the variants  can be  described by  the following
Hamiltonian:
\begin{multline}
{\cal  H  }  =  -\sum_{<ij>}^{NN}  \delta(\vec{S}_i,\vec{S}_j)+\lambda
\sum_{i,j=1}^N   \frac{|(\vec{S}_i   \cdot  \vec{r}_{ij})(\vec{S}_j   \cdot
\vec{r}_{ij})|}{|\vec{r}_{ij}|^3}\\
-H\sum_i^N(\vec{S}_i)^2+{\cal{H}}_{dis}
\end{multline}
\begin{figure}[h]
\begin{center}
\epsfig{file=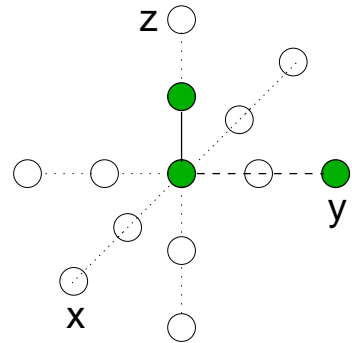,width=4.5cm,clip=}
\end{center}
\caption{\label{FIG1} (Color on  line) Schematic representation of the
nearest neighbors in a cubic lattice,  indicating the NN (continuous line) and
4N (dashed line) positions.}
\end{figure}

The  first contribution in the Hamiltonian 
is a  Potts  exchange term  between nearest neighbors (NN)  which
accounts  for the  bulk energy  favoring the  growth of  single phase
domains.   The  second is  the  long-range  dipolar interaction.   The
absolute  value  taken  in  the  dipolar term  guarantees  
the  rotational
invariance that is expected in the systems that we want to model.  The
third Hamiltonian term represents the interaction between the external
driving      field      $H$      and     the      order      parameter
$M=\sum_{i=1}^N(\vec{S}_i)^2$.    Note  that  the   vector  $\hat{0}$,
corresponding  to an  elementary domain  in the  austenite  phase, has
absolute value zero and thus does  not contribute to $M$. On the other
hand,  all  the three  variants  of  the  martensitic phase  are  here
represented by  vectors with modulus one.  Consequently,  $M/N$ can be
read as the normalized transformed fraction of the sample. 
Thus, the field $H$ drives
the  system  from  the  cubic  phase $m=0$  (when  $H=-\infty$)  to  a
multi-variant tetragonal phase  with $m=1$ (when $H=\infty$). It therefore
plays  the role of the  temperature for such  an ``athermal'' phase
transition.

The last term of the Hamiltonian accounts for the interaction with the
quenched disorder (impurities, dislocations, vacancies, etc.), 
which is always
present in real materials. Its precise form will be discussed later on
in this section.

As  explained in  the  introduction,  a first  version  of this  model
considered  only  nearest-neighbor  dipolar  interactions 
\cite{Cerruti2008}.   This
produced anisotropic  domains of the  different variants, but  not the
correct microstructures observed  in MT.  In the present  work we will
introduce higher order dipolar interactions. As will be seen, it is
enough to include the  interaction between fourth nearest 
neighbors (4N),
at a distance $2a$ from the reference elementary domain (see
Fig.~\ref{FIG1}).   It will be the interplay  between the interactions
to NN  and to 4N (which  are placed along the  same spatial direction)
that  generates the convenient  microstructures, in  a way  similar to
what happens in the ANNNI model \cite{Selke1988}.

Although  we  have  also   studied  the  second  and  third  
nearest-neighbors
interactions, we have found out that these  terms are not  crucial for the
microstructure formation  and do not  add any new  physics.  Therefore,
for simplicity in  most of the paper they  have been neglected, except
where indicated  otherwise.   Moreover,  we  have allowed  the  coefficients
$\lambda/\left|\vec{r}_{ij}\right|^3$  that control  the decay  of  the dipolar
interaction to  have more  freedom.  Since we are  truncating and
including  only two  terms, there  is no  need to  keep the coefficients  
as  in the
original model. We have  considered two general parameters $\lambda_1$
and $\lambda_2$  multiplying the dipolar  interaction to NN and  to 4N
nearest-neighbors. Thus, after these modification, our Hamiltonian reads:
\begin{multline}
{\cal  H  }  =-\sum_{<ij>}^{NN}  \delta(\vec{S}_i,\vec{S}_j)+\lambda_1
\sum_{ij}^{NN}|(\vec{S}_i    \cdot    \hat{r}_{ij})(\vec{S}_j    \cdot
\hat{r}_{ij})|\\    +\lambda_2   \sum_{ij}^{4N}    |(\vec{S}_i   \cdot
\hat{r}_{ij})(\vec{S}_j              \cdot              \hat{r}_{ij})|
-H\sum_i^N(\vec{S}_i)^2+{\cal{H}}_{dis},
\label{hamiltoniana}
\end{multline}
where the second sum runs over NN and the third only over the fourth 
nearest-neighbors.
The interaction  with quenched disorder has also been simplified from
what  was originally proposed  in Ref.~\onlinecite{Cerruti2008}.   Here, in
fact, we will not focus on the study of the influence of disorder, 
but we
need a certain  amount of it in order to favor nucleation of the
martensitic  phase  in all  the  equivalent  variants.   Thus we  have
considered:
\begin{equation}
{\cal{H}}_{dis}=-\sigma\sum_i^N\vec{g}_i\cdot\vec{S}_i,
\end{equation}
where  the  quenched random  fields  $\vec{g}_i$  are independent  and
identically distributed Gaussian random  variables, with zero mean and
unitary variance.   The parameter $\sigma$ models the  strength of the
disorder.   Note  that  after  the  simplifications,  the  Hamiltonian
parameters are only three: $\lambda_1$, $\lambda_2$, and $\sigma$.

For the study  of athermal behavior, apart from the Hamiltonian, one
must choose  a certain dynamics.   We chose to implement the extremal
update dynamics \cite{Cerruti2008}:  starting from a saturated
system  configuration,  corresponding  to  $m=0$  at  $H=-\infty$,  we
increase (decrease) the field by small steps $\Delta H$. At each field
step we  check the  contribution to the  energy of each  ${\vec S}_i$.
When a variable is found to  decrease the total energy by varying to a
new state,  we change  it to  the value that  gives the  most negative
energy change. In  this way we guarantee that the  system 
will reach the
same state  for the  same applied field  values, independently  of the
$\Delta   H$  value.    The  states reached are  equivalent   both
macroscopically  (same transformed  fraction $m$)  and microscopically
(same values of  $\vec{S}_i$ for all the $i$  values). The simulations
presented below were obtained by choosing $\Delta H=0.05$.

\section{Results}
\label{result}
\subsection{Ground state microstructures}
\label{groundstate}
Starting  from the  experimental observations from cubic  
to tetragonal
transitions,  our goal  would  be the  growth  of the  so-called  twin
variants.  These  correspond to  regions of alternating  $\hat{x}$ and
$\hat{y}$ (e.g.) domains separated by interfaces parallel to the $z$ axis and
forming  an  angle  of  $45^{\circ}$   with  the  $x$  and  $y$  axis.
Equivalent structures will  be the $\hat{x}$-$\hat{z}$ twins separated
by interfaces parallel  to the $y$ axis and  forming $45^{\circ}$ with
the  $x$ and  $z$  axis, and  with the $\hat{y}$-$\hat{z}$  twins separated  by
interfaces parallel to the $z$  axis and forming $45^{\circ}$ with the
$x$ and $y$ axis.

\begin{figure}[h]
\begin{center}
\epsfig{file=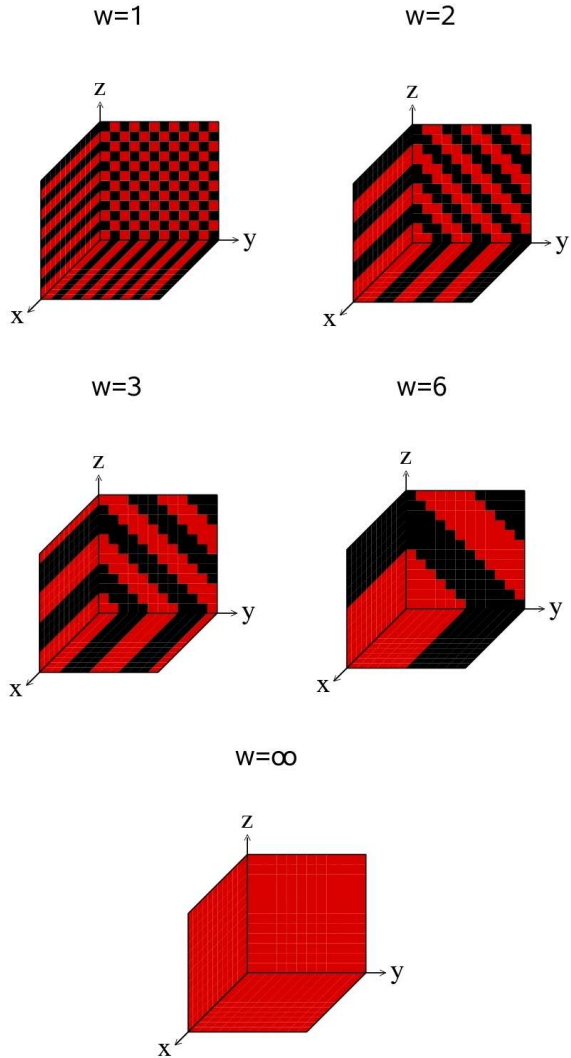,width=8cm,clip=}
\end{center}
\caption{\label{Fig2}  (Color online)  System configurations  of 
$\hat{y}-\hat{z}$
twins with  widths $w=1,2,3,6,\infty$, plotted for a  system with size
$L=12$.}
\end{figure}

In addition, we would  be interested in controlling the  width $w$ of such
twins. If one truncates the dipolar term to only first and second 
nearest-neighbors, 
it is  not possible to stabilize such  twin variants, except for
the case  with width  $w=1$, which in  fact corresponds  to domains
displaying    a    chessboard     structure    as    illustrated    in
Fig.~\ref{Fig2}(a).   Due  to  lattice geometry, third-nearest
neighbor interaction can not lead to  wider structures and it is only
by including 4N terms that  we can obtain a $w=2$ structure, which
is represented in Fig.~\ref{Fig2}(b).

In  order to  study the  range of  parameters in  which  the different
microstructures occur, we have  performed a comparison of the energies
of  the  different   configurations  shown  in  Fig.~\ref{Fig2},  with
$w=1,2,3,6,\infty$. The analytical  results presented here are general
for  any system  size provided  that the  proposed  configurations are
compatible with periodic boundary conditions.

The values of the energy as a function of the model parameters (in the
absence of  disorder) of each  of the configurations considered above
are given by:
\begin{eqnarray}
E_{1}&=&-N+N\lambda_2                   \nonumber                  \\
E_{2}&=&-{2}N+N\frac{1}{2}\lambda_1            \nonumber           \\
E_3&=&-\frac{7}{3}N+N\frac{2}{3}\lambda_1+N\frac{1}{3}\lambda_2
\nonumber       \\      E_6&=&\frac{47}{288}N(-4+\lambda_1+\lambda_2)+
\frac{97}{288}(-6+2\lambda_1+2\lambda_2)         \nonumber         \\
E_{\infty}&=&-3N+N\lambda_1+N\lambda_2.
\label{energie_lambda4}
\end{eqnarray}

\begin{figure}[h]
\begin{center}
\epsfig{file=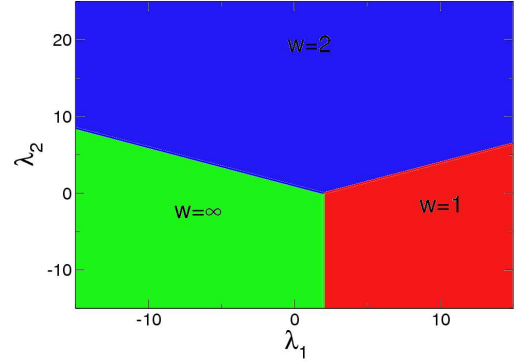,width=8cm,clip=}
\end{center}
\caption{\label{Fig3}   (Color   online)    Phase   diagram   in   the
$\lambda_1$-$\lambda_2$   plane.   The   solid  lines   represent  the
interfaces       between      the      phases       obtained      from
Eqs.\ref{energie_lambda4}.}
\end{figure}

From  such energy  functions one  can  obtain the  phase diagram  that
indicates  the boundaries  between  the structures  that minimize  the
energy. This is shown in Fig.\ref{Fig3}.

\begin{figure}[h]
\begin{center}
\epsfig{file=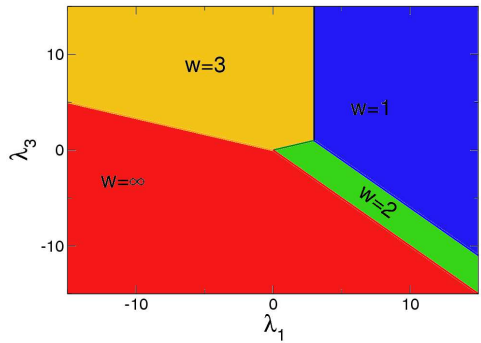,width=7.5cm,clip=}
\end{center}
\caption{\label{Fig4} (Color online) Phase     diagram     in      the     plane
$\lambda_1$-$\lambda_3$ for $\lambda_2=2$.}
\end{figure}

As can be seen, the  Eqs. \ref{energie_lambda4} allow the existence of
ground-state twinned structures  $w=1$, $w=2$, and $w=\infty$,  while no other
phases can exist. Although the  $w=2$ structure is good enough for our
objective of  studying kinetic constraints,  we have also  analyzed the
modifications one should perform on  the model in order to obtain 
ground-state twin structures with larger  widths.  The guess is that higher
order dipolar terms would lead  to the stabilization of  the wider
$w$  domains. To  confirm this  guess, let  us once again  consider  a $d=12$
system, and  add the  seventh nearest-neighbor  term (that is,  the neighbors  at a
distance  of  $3a$ from  the  reference  elementary domain).  The
energies of the various phases are now:
\begin{eqnarray}
E_{\infty}&=&-3N+N\lambda_1+N\lambda_2+N\lambda_3     \nonumber    \\
E_1&=&-N+N\lambda_2                    \nonumber                   \\
E_2&=&-{2}N+N\frac{1}{2}\lambda_1+\frac{N}{2}\lambda_3  \nonumber  \\
E_3&=&-\frac{7}{3}N+N\frac{2}{3}\lambda_1+N\frac{1}{3}\lambda_2.
\label{energie_lambda6}
\end{eqnarray}
An example of a projection of  the phase diagram is shown in Fig. \ref{Fig4},
where it  can be seen that the  $w=3$ phase is now  stable. The borders between     the     phases    are     calculated     with
Eqs.~\ref{energie_lambda6}.

For the remainder of the paper we will concentrate on the $\lambda_3=0$ case,
and in the region where
$w=2$.    Most   of  the   results   presented   will  correspond   to
$\lambda_1=10$ and $\lambda_2=20$.
 
\subsection{Kinetically obtained microstructures}
\label{simmicro}
\begin{figure}[h]
\begin{center}
\epsfig{file=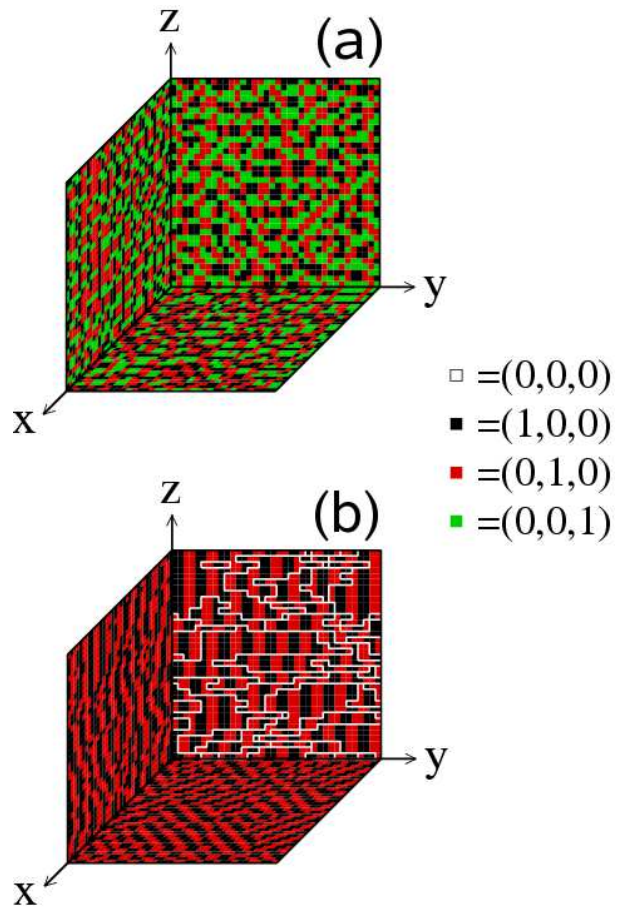,width=8.cm,clip=}
\end{center}
\caption{\label{Fig5} (Color  online) Examples of  two microstructures
  at  saturation for  $\lambda_1=10$, $\lambda_2=20$,  $\sigma=2.$ and
  $L=36$. In (a), the three variant saturation microstructure, in (b), the
  stressed configuration obtained by the application of a high external
  compression in the $z$ direction. White lines in the $yz$ planes are guides
  to the eye for the identification of the regions of twinned domains.}
\end{figure}

In Fig.\ref{Fig5}(a) we show an example of microstructures obtained by
the  numerical simulations  of the Hamiltonian (Eq.   \ref{hamiltoniana})
using athermal dynamics.   We represent only  three perpendicular
projections  of   our  3D  calculation.  This example thus only supplies   
qualitative
information.   In  section \ref{fourier_analysis}  we  will perform  a
Fourier analysis in order to gain some global information.

From the Figure \ref{Fig5}(a) it is  clear that some 
local structures of  the kind
$w=2$ arise, but that we  are very far from the ground-states computed
previously (see Fig.\ref{Fig2}). The  global structure is more complex and
displays mixing of  the three possible  twin domains.   Taking into
account that there are $3$ twin pairs and that each of them can occur in $2$
different  equivalent  orientations  and  that, given  our  underlying
discrete lattice, there are $4$ possible translations (for our $w=2$ case),
the number of different twin domains is $24$.

Another interesting example is given is Fig.~\ref{Fig5}(b). 
This figure corresponds to the saturation configuration obtained by
the application of a very high 
external compression in the $z$ direction, disfavoring
the associated $\hat{z}$ variant. 
In this case, the formation of diagonal $w=2$ domains
in the $x-y$ plane is already very clear at first sight, confirming the
importance of kinetic constraints arising due to the coexistence of the three
equivalent variants. It is interesting too to analyze the microstructure of 
the $x-z$ and $y-z$ planes. In this case, quite elongated
domains formed by vertical alternating stripes with periodicity $2$ arise, as
indicated in the figure.
This kind of domains have been observed experimentally in a iron-palladium
ferromagnetic shape-memory alloy, displaying a cubic to tetragonal martensitic
transition \cite{Cui2004}.

\subsection{Hysteresis}
\label{isteresi}
In this  section we  study the effect  of the  system size and  of the
Hamiltonian parameters  on the hysteresis  cycles. As will be  seen in
all of the  figures of this  section, the two  branches of the  loops are
strongly  asymmetric,  due to  the  physical  difference between  the
process of transformation from a one-variant phase to a three-variant
phase (ascending branch) and {\it vice-versa} in the descending case.

\begin{figure}[h]
\begin{center}
\epsfig{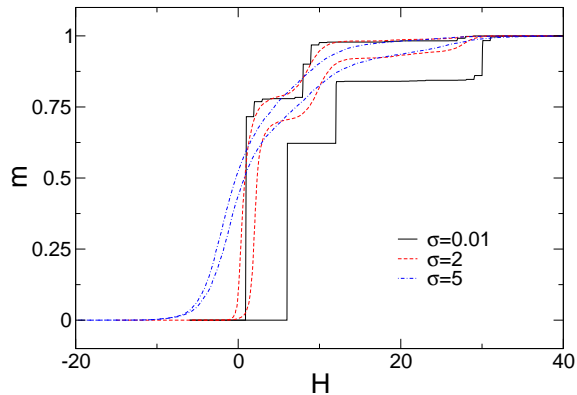}
\end{center}
\caption{\label{Fig6}    (Color   online)   Hysteresis    cycles   for
$\sigma=0.01;2;5$,  for   system  parameters  $L=24$,  $\lambda_1=10$,
 and $\lambda_2=20$.}
\end{figure}

In  Fig.   \ref{Fig6},  we  show  some examples  of  hysteresis  loops
obtained by  varying the  amount of disorder  $\sigma$.  The loops  have a
tendency to display plateaus,  which are smooth when disorder
increases.   An additional  effect  of  disorder  is  also to  tilt  the
hysteresis loops.

\begin{figure}[h]
\begin{center}
\epsfig{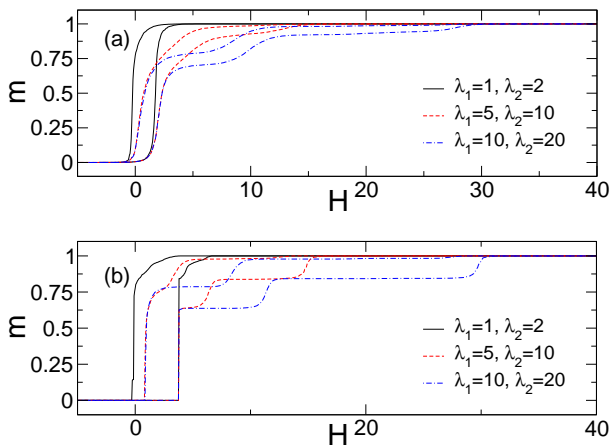}
\end{center}
\caption{\label{Fig7}    (Color   online)   Hysteresis    cycles   for
$(\lambda_1,\lambda_2)=(1,2);(5,10);(10,20)$,  for  system  parameters
$L=24$, and $\sigma=2$ (a) and $\sigma=0.5$ (b).}
\end{figure}

In Figs. \ref{Fig7}, we present  examples of the effect of the dipolar
parameters $\lambda_1$ and $\lambda_2$. Within the region of stability
of the $w=2$ phase, the plateaus are always present, provided that 
disorder is not too high.

\begin{figure}[h]
\begin{center}
\epsfig{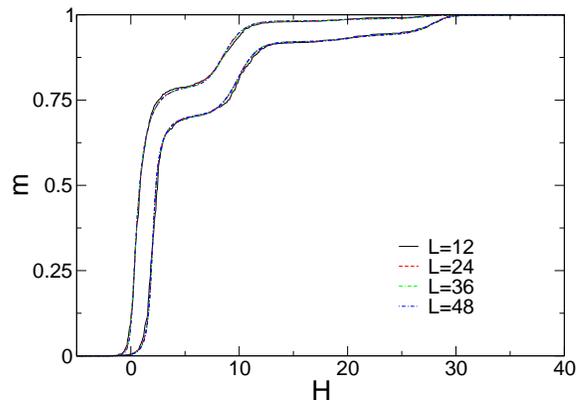}
\end{center}
\caption{\label{Fig8} (Color  online) Hysteresis loops  for $L=12,24,36,48$,
for system parameters $\lambda_1=10$, $\lambda_2=20$ and $\sigma=2$.}
\end{figure}

Finally, in Fig. \ref{Fig8} we study  the effect of system size on
the hysteresis loops. The loops  do not show a strong size dependence,
and  the  plateaus are  clearly  not a  finite-size  effect.  In  the
following  we   will  restrict  our  simulations   to  the  affordable
intermediate size $L=36$.

All the hysteresis  cycles that we have shown  in this section display
qualitatively the same behavior,  consisting of a certain number (2-3)
of plateaus, where the transformed  fraction remains constant in a range
of $H$  values. This feature  suggests that the transformation  of the
$\hat{0}$  phase has a  tendency to  split into  steps, the  first
involving the transformation of more than 50\% of the lattice.  In
the  next sections we  will try  to elucidate  the physical
reasons behind this tendency.

\subsection{Real space analysis of the transition dynamics}
\label{real_space_analysis}

\begin{figure}[h]
\begin{center}
\epsfig{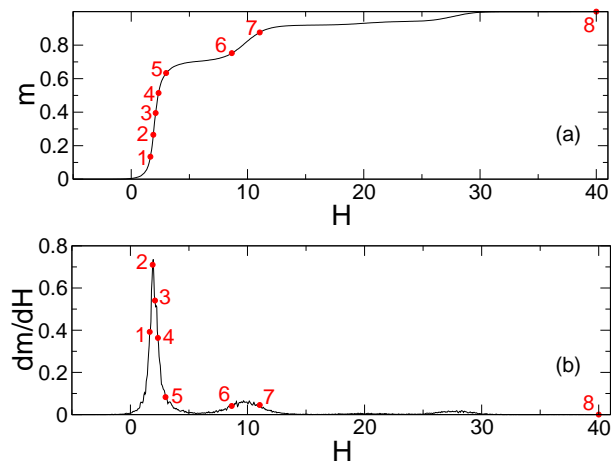}
\end{center}
\caption{\label{Fig9} (Color online)  Hysteresis ascendant loop branch
  (a) and  (b)field derivative of the  normalized transformed fraction
  $dm/dH$ as a function  of the field, for $L=36$, $\lambda_1=10$,
  $\lambda_2=20$, and $\sigma=2$.  The points marked
  with numbers $1,2,3,4,5,6,7,8$  correspond to snapshots  in Fig.~\ref{Fig10}
  and have $m=0.125;0.25;0.375;0.5;0.625;0.75;0.875;1$ respectively.}
\end{figure}

In order  to investigate the nature  of the intermediate  steps of the
hysteresis cycles we can
analyze the system  configurations in real space at  various points of
the  loops,  as indicated  in  Fig.\ref{Fig9}(a).   We  also show  the
derivative  $dm/dH$   which  better reflects  the   activity  of  the
transition and reveals the importance of the first transformation step
compared to the others. These kinds of activity curves can be compared 
with the experimental data from calorimetry or acoustic emission activity.

\begin{figure*}[t]
\begin{center}
\epsfig{file=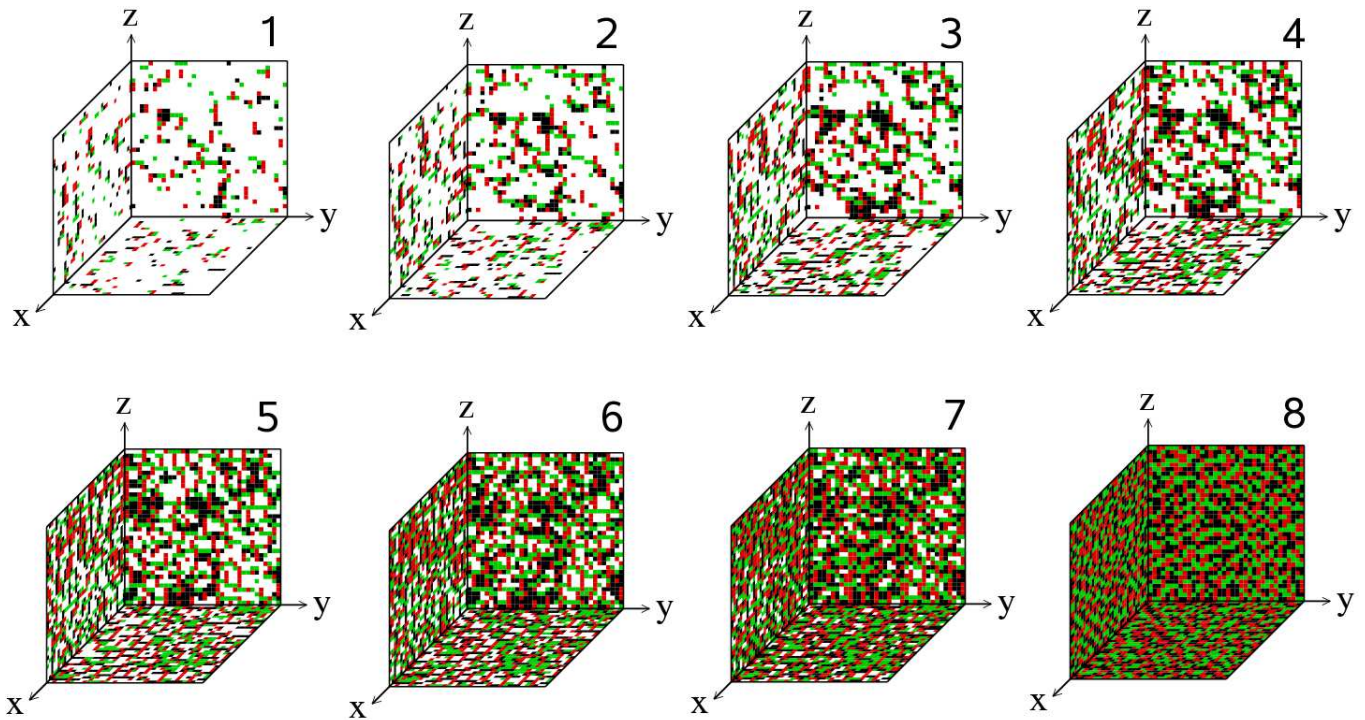,width=18cm,clip=}
\end{center}
\caption{\label{Fig10}   (Color  online)   Snapshots  of   real  space
configurations corresponding  to the  field loop points  indicated in
Fig. \ref{Fig9}.}
\end{figure*}

The microstructures  on each indicated  point of the  ascending branch
are  shown in Fig.~\ref{Fig10}. A first  inspection  reveals  that the  
transition
basically proceeds by nucleation  events.  There  is some  growth, but
almost no coarsening since the  possibilities for two domains to match
are  small.  One  can   also  see  that  re-transformation  events  are
common. The observed step in the hysteresis loop is clearly associated
with the fact  that martensite domains percolate in  the sample. At this
point  the system needs an  extra increase  of the  driving  force to
proceed. The subsequent domains will grow surrounded by a complex boundary
of the martensitic phase. Therefore, the growth dynamics will be very different
from the nucleation of the first domains in a cubic phase environment.
In  the second part,  there are many  re-transformation events
until the  final saturated microstructure  is obtained.  After  a more
careful study of the snapshots one can infer that in the first step of
the transformation,  corresponding to the  snapshots from $1$  to $6$,
the system chooses to develop oblate (disk-like) domains. For instance,
one can see in snapshot number 3 that the $\hat{x}$-phase (black)
tends to be distributed in extended regions in the $y-z$ plane, whereas
it shows elongated domains perpendicular  to the $x$-axis on the $x-y$
and $x-z$  planes.  Such structures correspond to  the structures that
optimize the NN dipolar  interaction term ($\lambda_1$), as we studied
in a  preliminary work \cite{Cerruti2008}.  In a second  step, when the
external  field   is  comparable  to   the  value  of   the  parameter
$\lambda_2$, the system  is able to develop the  $w=2$ structures with
the  desired  orientation, and  one  starts  to  see regions  of  twin
domains with  the $45^\circ$  orientation on the  different planes.
Of course, they  are very much influenced  by the domains that
have  already  grown.   This  interpretation  can even be argued  by
confronting  the  curves  for  different  values  of  $\lambda_1$  and
$\lambda_2$   in  Fig.    \ref{Fig7}.   For   a   better  quantitative
understanding of this effect, it is convenient to perform the analysis
of the microstructures in Fourier space.

\subsection{Fourier analysis}
\label{fourier_analysis}
A quite natural  way to analyze the system  microstructure is to study
the Fourier  Transform (FT)  of the  density function. In fact,  as it  is well
known, the  square modulus of  this quantity represents  the intensity
that can  be obtained  from scattering experiments.   In our  case, at
saturation  we  have the  coexistence  of  three variants,  $\hat{x}$,
$\hat{y}$, and $\hat{z}$, and thus it is useful to separately consider 
three         density         matrices        $\rho_{\hat{x}}(x,y,z)$,
$\rho_{\hat{y}}(x,y,z)$, and $\rho_{\hat{z}}(x,y,z)$. In analogy we can
define the density matrix of the $\hat{0}$ phase $\rho_{\hat{0}}(x,y,z)$,
and transform the four matrices individually. Thus each variant $\alpha$
($\alpha=\hat{0},\hat{x},\hat{y},\hat{z}$) is associated with a scattered
intensity:
$$I_{\alpha}(K_x,K_y,K_z)\propto\left|\sum_{b,c,d=1}^L\rho_{\alpha}(b,c,d)e^{i\frac{2\pi}{L}(K_xb+K_yc+K_zd)}\right|^2,$$
where $\vec{r}=x\hat{i}+y\hat{j}+z\hat{k}=ba\hat{i}+ca\hat{j}+da\hat{k}$ is
the generic position vector, and $K_x,K_y,K_z=0,L-1$.
 The elements of these four  density matrices can
take the  values $1$ or $0$, representing,  respectively, 
the occupation
or the vacancy of  a lattice site by a variant of  a given color.  The
presence of  disorder can locally favors one  or the other variant,
but since the  disorder is randomly extracted, we  expect variants
behavior to be  equivalent on average.  We thus  restrict ourselves to
studying  just one  variant, e.g.   the  $\hat{x}$ one,  without loss  of
generality.   In the simulations we  have verified that the  color
equivalence  is  effectively  fulfilled.   In the  following  we  will
address only this case.

\begin{figure}[t]
\begin{center}
\epsfig{file=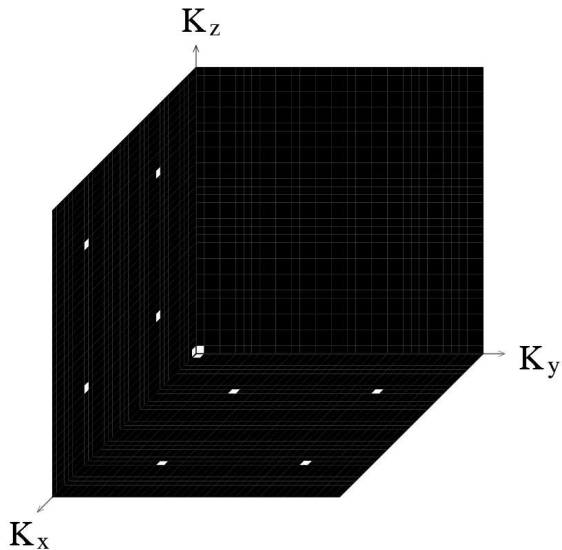,width=7.5cm,clip=}
\end{center}
\caption{\label{Fig11} Positions of the peaks of the scattered  intensity 
$I_{\hat{x}}$ for the
  ideal  case  with  $w=2$,  for  a system  with  $L=36$  and  without
  disorder. The  configuration is obtained  as a superposition  of the
  six possible orientations of the $w=2$ stripes.}
\end{figure}

In Fig.\ref{Fig11} we show the  position of the peaks of the 
squared Fourier Transform of the density of the $\hat  x$
variant in the ideal
case  $w=2$.  We  observe  that  Bragg peaks  occur  at the  position
$\vec{K}=(L/4,L/4,0)$ and in symmetric  equivalent positions.  
In fact, if one
performs  the  Fourier transform  of  the  density  of $\hat  x$-phase
corresponding,  for instance,  to the  snapshot in  Fig.\ref{Fig2}, one
obtains   only  two   of  the   eight  Bragg   reflections   shown  in
Fig.\ref{Fig11}.   Nevertheless,  for  symmetry  reasons  one  has  to
consider the  two possible $45^\circ$  orientations and also  that the
phase $\hat  x$ would  be present not only in the $\hat{x}-\hat{y}$ twin,
but also in the $\hat{x}-\hat{z}$ twin. 
Translations of the
configuration with the same domain orientation give the same
contribution. Therefore, we
shall  expect the $8$  Bragg reflections  shown in  Fig.\ref{Fig11} with
equal  intensity 
$I_{\hat{x}}(L/4,L/4,0)=1/6(N^2/8)$,
where the contribution $N^2/8$ is given by the intensity of the two secondary
peaks that arise for each of
the four ideal configurations where the variant $\hat{x}$ is present, and the
factor $1/6$ is due to the average over all the six possible ideal
configurations.  The  peak  at  the  origin stands  
for  the
fraction of  the system  which is  covered by  the $\hat x$ phase, and thus
its intensity is given by $I_{\hat{x}}(0,0,0)=4/6(N^2/4)$, where again the factor $4/6$ is
due to the average over the six possibilities. 
In a real experimental case, the widths of the twin domains will
be much bigger than the $w=2$ lattice spacing and, therefore the growth of
the martensitic tetragonal domains will be revealed by satellite peaks
occurring  close to  the  cubic Bragg  reflections  in the  directions
(1,1,0) (and equivalent ones).

\begin{figure}[h]
\begin{center}
\epsfig{file=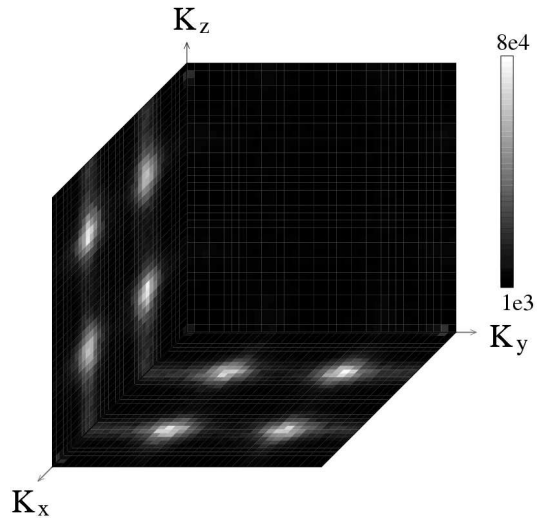,width=7.cm,clip=}
\end{center}
\caption{\label{Fig12}  Scattered   intensity  $I_{\hat{x}}$  (linear  scale):
Fourier  transform of  the  $x$  particle density  for  a sample  with
$L=36$,  $\sigma=2$  $\lambda_1=10$,  and $\lambda_2=20$.   Data  are
averaged over $100$ realizations of the disorder.  The $\vec{K}=(0,0,0)$ peak
has been suppressed in order to enhance the contrast.  }
\end{figure}

In   Fig.~\ref{Fig12}   we   show   the  Fourier   analysis   of   the
microstructures obtained after saturating  the system by the numerical
simulations.   The  intensity is averaged  over  $100$
realizations of disorder.  We see  that the peaks appear at the
positions expected from the ground-state structures, but the widths of
the peaks are quite large.

\begin{figure*}[h]
\begin{center}
\epsfig{file=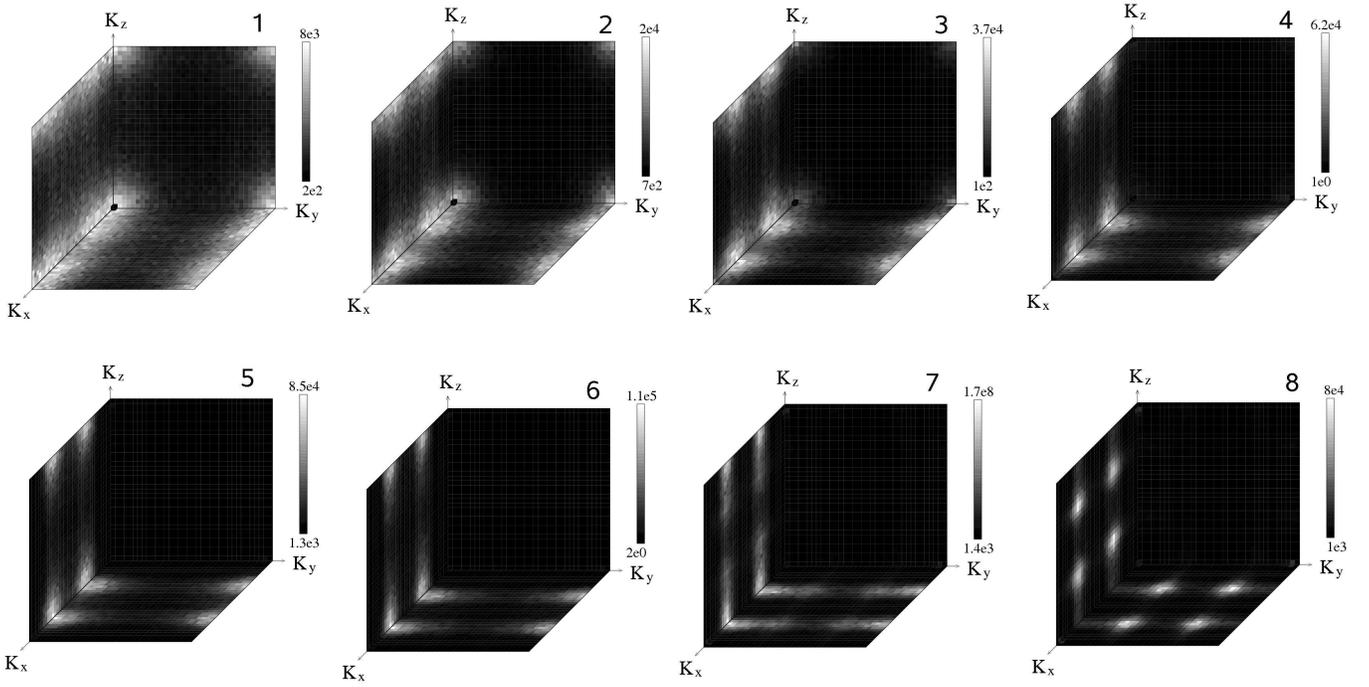,width=18cm,clip=}
\end{center}
\caption{\label{Fig13}  Scattered intensity $I_{\hat{x}}(\vec{K})$  (linear   scale)   for
$\sigma=2$, $\lambda_1=10$, $\lambda_2=20$, and $L=36$. Data
are averaged over $100$  disorder realizations.  The snapshots correspond to
the points indicated in Fig.~\ref{Fig9}:
1) $m=0.125$, 2) $m=0.25$, 3)  $m=0.375$, 4) $m=0.5$, 5) $m=0.625$, 6)
$m=0.75$, 7) $m=0.875$, and 8) $m=1$. The $\vec{K}=(0,0,0)$ peaks
have been suppressed in order to enhance the contrast. }
\end{figure*}

In analogy  with the study performed  in real space, we  can study the
evolution of  the Fourier transform along the  hysteresis cycles. This
is  presented in  Fig.~\ref{Fig13}.  As  we have already
noticed, the system evolution begins  with a tendency to form and grow
disk-like domains  which correspond  to peaks  at  $\vec{K}=(L/4,0,0)$ (and
equivalent). After the first plateau the intensity is transferred
to the final positions at $\vec{K}=(L/4,L/4,0)$ (and equivalent).

\begin{figure}[h]
\begin{center}
\epsfig{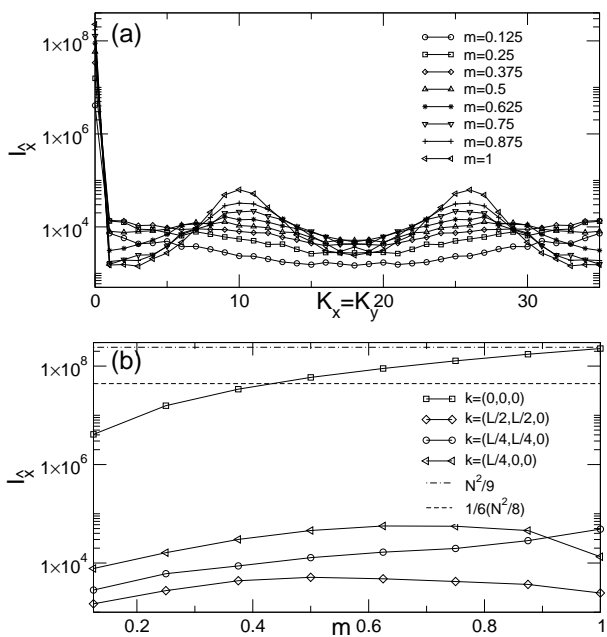}
\end{center}
\caption{\label{intensita}  (a) Intensity  
$I_{\hat{x}}$ for $m=0.125;0.25;0.375;0.5;0.675;0.75;0.875;1$  for  the
ascendant  branch on  the $K_x=K_y$  diagonal, with $K_z=0$. (b) Values of 
the scattered intensity as a function of the reduced transformed fraction, 
for $\vec{K}=(0,0,0)$, $\vec{K}=(L/2,L/2,0)$, $\vec{K}=(L/4,L/4,0)$, and
$\vec{K}=(L/4,0,0)$. Dot-dashed and dashed line represent the theoretical 
values of saturation for ideal configurations of the $\vec{K}=(0,0,0)$ and the
$\vec{K}=(L/4,L/4,0)$ peaks.  The
values  are  averaged over  $100$  disorder  configurations, for  system
parameters  $\sigma=2$, $\rho=0$, $\lambda_1=10$,  $\lambda_2=20$, and
$L=36$.}
\end{figure}

In  order to  gain a  quantitative insight  into the  dynamics  of the
transition, we have analyzed the evolution of the scattered intensity
corresponding to some significant vectors of
the Fourier space. In Fig. \ref{intensita}(a), we present the evolution of the
diagonal $K_x=K_y$ with $K_z=0$. As can be seen, the peak is slightly
shifted from the ideal position $(L/4,L/4,0)$ (and equivalent). In particular,
the position is closer to the origin, meaning that the formation of some
thinner structure is affecting the average. This effect is due to the presence
of the disorder and of the dynamic constraints. In Fig.\ref{intensita}(b), we
present the evolution of the intensity as a
function of the transformed fraction $m$. We choose $m$ and not $H$ as the
independent variable since, as can be seen in Figs.~\ref{Fig6}, 
~\ref{Fig7} and ~\ref{Fig8}, the jumps in the
hysteresis cycles are very sharp. As can be seen, the
$\vec{K}=(0,0,0)$ and the $\vec{K}=(L/4,L/4,0)$ values grow monotonically with
$m$ as one could expect, and the intensity at $\vec{K}=(L/4,L/4,0)$ and 
$\vec{K}=(L/2,L/2,0)$ tends to disappear at saturation. The scattered 
intensity in this latter position is a
measure of the disorder in the system, since in the ideal
configuration the corresponding value of $I_{\hat{x}}$ is zero. In the figure we
confront the $\vec{K}=(0,0,0)$ and the $\vec{K}=(L/4,L/4,0)$ values with the
related theoretical values at saturation. The value of the $\vec{K}=(0,0,0)$
peak for $m=1$ is in perfect agreement with the expected one, while the 
$\vec{K}=(L/4,L/4,0)$, even if increasing, as it should do in the presence of
$w=2$ structures, keeps far from the ideal value. Furthermore, the predominance of
the $w=2$ structures arises only in the last part of the hysteresis branch,
where the reduced transformed fraction goes from $m=0.875$ to $m=1$. These
effects are due to the presence of disorder and of the kinetic constraints
associated with the presence of
many competing growing domains. 

\section{Conclusions}
\label{Conclusions}

We have presented a model for the statistical study of microstructures
in martensitic phase transitions. We have  focused on the cubic to tetragonal
case.    The  model  is   not  aimed at exactly   reproducing  the
microstructure details but only to provide a statistical picture of the
different  phenomena that may  occur during  the transition  and that
will constrain the final martensitic state.

Our model allows some  dynamical aspects of the MT that occur
due to  the existence  of kinetic constraints to be studied.   Among others,  we have
shown the existence of domain re-transformation, the fact that the
transition  may proceed through  several steps,  and the  existence of
intermediate structures that minimize only the short-range term of the
dipolar   interaction.   The   Fourier  transform   of   the  obtained
microstructures  has  enabled  a quantitative  analysis  of  such
phenomena   after  taking   averages  over   disorder  configurations.
Although some of  the observed phenomena may not be totally realistic
due  to  the  cut-off  that   we  have  introduced  in  the  long-range
interaction, we  have shown that the  model is good  enough to exhibit
kinetic constraints. The possibility for extending the interactions to
real  long-range interactions is still  open and,  although it  will  require much
longer computing times, should not be excluded for a future work.

In our opinion, the presented results quite
clearly justify the importance of a statistical approach to the martensitic phase
transition. 
We finally  propose that, besides  the real imaging  experiments, many
more ``in situ'' scattering experiments  will be desirable in order to
gain understanding  into the  dynamics of microstructure  formation in
martensites.

\acknowledgments The authors  acknowledge fruitful discussions with A.
Planes. B.C. would also like to thank P. Moretti for very helpful 
conversations.  This work has  received financial support from 
CICyT (Spain), project MAT2007-61200 and  CIRIT (Catalonia), 
project 2005SGR00969 and
Marie Curie RTN MULTIMAT (EU), Contract No. MRTN-CT-2004-5052226.


\begin{thebibliography}{18}
\expandafter\ifx\csname natexlab\endcsname\relax\def\natexlab#1{#1}\fi
\expandafter\ifx\csname bibnamefont\endcsname\relax
  \def\bibnamefont#1{#1}\fi
\expandafter\ifx\csname bibfnamefont\endcsname\relax
  \def\bibfnamefont#1{#1}\fi
\expandafter\ifx\csname citenamefont\endcsname\relax
  \def\citenamefont#1{#1}\fi
\expandafter\ifx\csname url\endcsname\relax
  \def\url#1{\texttt{#1}}\fi
\expandafter\ifx\csname urlprefix\endcsname\relax\def\urlprefix{URL }\fi
\providecommand{\bibinfo}[2]{#2}
\providecommand{\eprint}[2][]{\url{#2}}

\bibitem[{\citenamefont{H.Zhang et~al.}(2008)\citenamefont{H.Zhang,
  E.K.H.Salje, D.Schryvers, and B.Bartova}}]{Zhang2008}
\bibinfo{author}{\bibnamefont{H.Zhang}},
  \bibinfo{author}{\bibnamefont{E.K.H.Salje}},
  \bibinfo{author}{\bibnamefont{D.Schryvers}}, \bibnamefont{and}
  \bibinfo{author}{\bibnamefont{B.Bartova}}, \bibinfo{journal}{J.
  Phys.:Condens. Matter} \textbf{\bibinfo{volume}{20}}, \bibinfo{pages}{055220}
  (\bibinfo{year}{2008}).

\bibitem[{\citenamefont{J.Cui et~al.}(2004)\citenamefont{J.Cui, T.W.Shield, and
  R.D.James}}]{Cui2004}
\bibinfo{author}{\bibnamefont{J.Cui}},
  \bibinfo{author}{\bibnamefont{T.W.Shield}}, \bibnamefont{and}
  \bibinfo{author}{\bibnamefont{R.D.James}}, \bibinfo{journal}{Acta Materialia}
  \textbf{\bibinfo{volume}{52}}, \bibinfo{pages}{35} (\bibinfo{year}{2004}).

\bibitem[{\citenamefont{K.Bhattacharya}(2003)}]{Bhattacharya2003}
\bibinfo{author}{\bibnamefont{K.Bhattacharya}},
  \emph{\bibinfo{title}{Microstructure of martensite}}
  (\bibinfo{publisher}{Oxford University Press}, \bibinfo{year}{2003}).

\bibitem[{\citenamefont{Toninelli et~al.}(2006)\citenamefont{Toninelli, Biroli,
  and Fisher}}]{Toninelli2006}
\bibinfo{author}{\bibfnamefont{C.}~\bibnamefont{Toninelli}},
  \bibinfo{author}{\bibfnamefont{G.}~\bibnamefont{Biroli}}, \bibnamefont{and}
  \bibinfo{author}{\bibfnamefont{D.~S.} \bibnamefont{Fisher}},
  \bibinfo{journal}{Phys.\ Rev.\ Lett.} \textbf{\bibinfo{volume}{96}},
  \bibinfo{pages}{035702} (\bibinfo{year}{2006}).

\bibitem[{\citenamefont{Toninelli and Biroli}(2006)}]{Toninelli2006b}
\bibinfo{author}{\bibfnamefont{C.}~\bibnamefont{Toninelli}} \bibnamefont{and}
  \bibinfo{author}{\bibfnamefont{G.}~\bibnamefont{Biroli}},
  \bibinfo{journal}{cond-mat/060380v1}  (\bibinfo{year}{2006}).

\bibitem[{\citenamefont{A.Hubert and R.Schaefer}(1998)}]{schaefer}
\bibinfo{author}{\bibnamefont{A.Hubert}} \bibnamefont{and}
  \bibinfo{author}{\bibnamefont{R.Schaefer}}, \emph{\bibinfo{title}{Magnetic
  domains}} (\bibinfo{publisher}{Springer}, \bibinfo{address}{New York},
  \bibinfo{year}{1998}).

\bibitem[{\citenamefont{G.Durin and S.Zapperi}(2006)}]{barkhausen}
\bibinfo{author}{\bibnamefont{G.Durin}} \bibnamefont{and}
  \bibinfo{author}{\bibnamefont{S.Zapperi}}, \bibinfo{journal}{"The Science of
  Hysteresis", G. Bertotti and I. Mayergoyz eds, Elsevier, Amsterdam}
  \textbf{\bibinfo{volume}{II}}, \bibinfo{pages}{181} (\bibinfo{year}{2006}).

\bibitem[{\citenamefont{Tendeloo et~al.}(1997)\citenamefont{Tendeloo,
  Meulenaere, and Schryvers}}]{Vantendeloo1997}
\bibinfo{author}{\bibfnamefont{G.~V.} \bibnamefont{Tendeloo}},
  \bibinfo{author}{\bibfnamefont{P.~D.} \bibnamefont{Meulenaere}},
  \bibnamefont{and}
  \bibinfo{author}{\bibfnamefont{D.}~\bibnamefont{Schryvers}},
  \bibinfo{journal}{Physica D} \textbf{\bibinfo{volume}{107}},
  \bibinfo{pages}{401} (\bibinfo{year}{1997}).

\bibitem[{\citenamefont{M.Mitsuka et~al.}(2006)\citenamefont{M.Mitsuka, T.Ohba,
  T.Fukuda, T.Kakeshita, and M.Tanaka}}]{Mitsuka2006}
\bibinfo{author}{\bibnamefont{M.Mitsuka}},
  \bibinfo{author}{\bibnamefont{T.Ohba}},
  \bibinfo{author}{\bibnamefont{T.Fukuda}},
  \bibinfo{author}{\bibnamefont{T.Kakeshita}}, \bibnamefont{and}
  \bibinfo{author}{\bibnamefont{M.Tanaka}}, \bibinfo{journal}{Mat. Sci. Eng. A}
  \textbf{\bibinfo{volume}{438-440}} (\bibinfo{year}{2006}).

\bibitem[{\citenamefont{Y.Wang and A.G.Khachaturyan}(1997)}]{Wang1997}
\bibinfo{author}{\bibnamefont{Y.Wang}} \bibnamefont{and}
  \bibinfo{author}{\bibnamefont{A.G.Khachaturyan}}, \bibinfo{journal}{Acta
  mater.} \textbf{\bibinfo{volume}{45}}, \bibinfo{pages}{759}
  (\bibinfo{year}{1997}).

\bibitem[{\citenamefont{T.Ichitsubo et~al.}(2000)\citenamefont{T.Ichitsubo,
  K.Tanaka, M.Koiwa, and Y.Yamazaki}}]{Ichitsubo2000}
\bibinfo{author}{\bibnamefont{T.Ichitsubo}},
  \bibinfo{author}{\bibnamefont{K.Tanaka}},
  \bibinfo{author}{\bibnamefont{M.Koiwa}}, \bibnamefont{and}
  \bibinfo{author}{\bibnamefont{Y.Yamazaki}}, \bibinfo{journal}{Phys. Rev. B}
  \textbf{\bibinfo{volume}{62}}, \bibinfo{pages}{5435} (\bibinfo{year}{2000}).

\bibitem[{\citenamefont{R.Ahluwalia and
  G.Ananthakrishna}(2001)}]{Ahluwalia2001}
\bibinfo{author}{\bibnamefont{R.Ahluwalia}} \bibnamefont{and}
  \bibinfo{author}{\bibnamefont{G.Ananthakrishna}}, \bibinfo{journal}{Phys.
  Rev. Lett.} \textbf{\bibinfo{volume}{86}}, \bibinfo{pages}{4076}
  (\bibinfo{year}{2001}).

\bibitem[{\citenamefont{Jacobs et~al.}(2003)\citenamefont{Jacobs, Curnoe, and
  Desai}}]{Jacobs2003}
\bibinfo{author}{\bibfnamefont{A.E.}~\bibnamefont{Jacobs}},
  \bibinfo{author}{\bibfnamefont{S.H.}~\bibnamefont{Curnoe}}, \bibnamefont{and}
  \bibinfo{author}{\bibfnamefont{R.C.}~\bibnamefont{Desai}},
  \bibinfo{journal}{Phys. Rev. B} \textbf{\bibinfo{volume}{68}},
  \bibinfo{pages}{224104} (\bibinfo{year}{2003}).

\bibitem[{\citenamefont{D.M.Hatch et~al.}(2003)\citenamefont{D.M.Hatch,
  T.Lookman, A.Saxena, and S.R.Shenoy}}]{Hatch2003}
\bibinfo{author}{\bibnamefont{D.M.Hatch}},
  \bibinfo{author}{\bibnamefont{T.Lookman}},
  \bibinfo{author}{\bibnamefont{A.Saxena}}, \bibnamefont{and}
  \bibinfo{author}{\bibnamefont{S.R.Shenoy}}, \bibinfo{journal}{Phys. Rev. B}
  \textbf{\bibinfo{volume}{68}}, \bibinfo{pages}{104105}
  (\bibinfo{year}{2003}).

\bibitem[{\citenamefont{T.Lookman et~al.}(2003)\citenamefont{T.Lookman,
  S.R.Shenoy, K.O.Rasmussen, A.Saxena, and A.R.Bishop}}]{Lookman2003}
\bibinfo{author}{\bibnamefont{T.Lookman}},
  \bibinfo{author}{\bibnamefont{S.R.Shenoy}},
  \bibinfo{author}{\bibnamefont{K.O.Rasmussen}},
  \bibinfo{author}{\bibnamefont{A.Saxena}}, \bibnamefont{and}
  \bibinfo{author}{\bibnamefont{A.R.Bishop}}, \bibinfo{journal}{Phys.\ Rev.\ B}
  \textbf{\bibinfo{volume}{67}}, \bibinfo{pages}{024114}
  (\bibinfo{year}{2003}).

\bibitem[{\citenamefont{K.O.Rasmussen et~al.}(2001)\citenamefont{K.O.Rasmussen,
  T.Lookman, A.Saxena, A.R.Bishop, R.C.Albers, and S.R.Shenoy}}]{Rasmussen2001}
\bibinfo{author}{\bibnamefont{K.O.Rasmussen}},
  \bibinfo{author}{\bibnamefont{T.Lookman}},
  \bibinfo{author}{\bibnamefont{A.Saxena}},
  \bibinfo{author}{\bibnamefont{A.R.Bishop}},
  \bibinfo{author}{\bibnamefont{R.C.Albers}}, \bibnamefont{and}
  \bibinfo{author}{\bibnamefont{S.R.Shenoy}}, \bibinfo{journal}{Phys.\ Rev.\
  Lett.} \textbf{\bibinfo{volume}{87}}, \bibinfo{pages}{055704}
  (\bibinfo{year}{2001}).

\bibitem[{\citenamefont{B.Cerruti and E.Vives}(2008)}]{Cerruti2008}
\bibinfo{author}{\bibnamefont{B.Cerruti}} \bibnamefont{and}
  \bibinfo{author}{\bibnamefont{E.Vives}}, \bibinfo{journal}{Phys. Rev. B}
  \textbf{\bibinfo{volume}{77}}, \bibinfo{pages}{064114}
  (\bibinfo{year}{2008}).

\bibitem[{\citenamefont{W.Selke}(1988)}]{Selke1988}
\bibinfo{author}{\bibnamefont{W.Selke}}, \bibinfo{journal}{Phys. Rep.}
  \textbf{\bibinfo{volume}{170}}, \bibinfo{pages}{213} (\bibinfo{year}{1988}).

\end{thebibliography}

\end{document}